\documentclass[onecolumn]{elsart}
\usepackage{amssymb,amsmath,graphicx}
\journal{Physica E}

\begin{document}

\begin{frontmatter}

\title{Growth and optical properties of self-assembled InGaAs Quantum Posts}

\author[materials]{H. J. Krenner\corauthref{cor}}
\author[iowa]{C. Pryor}
\author[materials]{J. He}
\author[materials]{J. P. Zhang}
\author[materials]{Y. Wu}
\author[physics]{C. M. Morris}
\author[physics]{M. S. Sherwin}
\author[materials,ece]{P. M. Petroff}
\corauth[cor]{krenner@engineering.ucsb.edu}
\address[materials]{Materials Department, UC Santa Barbara, Santa Barbara CA 93106, USA} 
\address[iowa]{Department of Physics and Astronomy, University of Iowa, Iowa City, IA 52242, USA} 
\address[physics]{Physics Department, UC Santa Barbara, Santa Barbara CA 93106, USA}
\address[ece]{Department of Electrical and Computer Engineering, UC Santa Barbara, Santa Barbara CA 93106, USA}


\begin{abstract}
We demonstrate a method to grow height controlled, dislocation-free $\mathrm{InGaAs}$ quantum posts (QPs) on GaAs by molecular beam epitaxy (MBE) which is confirmed by structural investigations. The optical properties are compared to realistic 8-band $k\cdot p$ calculations of the electronic structure which fully account for strain and the structural properties of the QP. Using QPs embedded in n-i-p junctions we find wide range tunability of the interband spectrum and giant static dipole moments.
\end{abstract}

\end{frontmatter}

\maketitle

Self-assembled Quantum Dots (QDs) have attracted widespread interest due to their large potential for application in novel devices for optoelectronics and quantum information processing. Their growth by molecular beam epitaxy (MBE) has been perfected over the last decade. This strain-driven self-assembly technique was successfully transferred from III-V to group IV and II-VI compounds demonstrating their large application potential and versatility. Due to the relative ease of growth and well established integration in semiconductor devices, the self-assembled QDs provide a great advantage over other types of nanostructures. In particular, realizations of Quantum Wires with two-dimensional carrier confinement have remained relatively unstudied due to the more complex fabrication.\\
Here we present a growth technique for height-controlled $\mathrm{InGaAs}$ Quantum Posts (QPs) using MBE. We study their structural and optical properties along with calculations of the electronic structure. Furthermore, we find  large Stark-shifts which directly reflect the QP height.


\section{Growth and structural characterization}
\begin{figure*}[hbt]
	\centering
		\includegraphics[width=.750\textwidth]{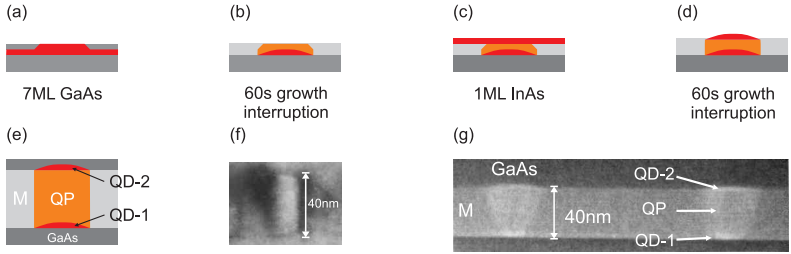}
		\caption{(a-d) The growth sequence of a QP. (e-f) Schematic of an embedded QP and X-TEM images.\label{fig1}}
\end{figure*}
The samples were grown by solid source MBE on semi-insulating GaAs (100) substrates. For the QP growth the substrate temperature was lowered to $530\mathrm{^oC}$ to deposit 2 monolayers (MLs) of InAs which form a layer of seed QDs acting as nucleation centers. The QPs are formed by  multiple repetitions of a deposition sequence shown schematically in Fig.\ref{fig1} together with cross-sectional transmission electron micrographs. Our deposition sequence consists of two steps of 7 GaAs MLs (a) and 1 ML of InAs (c), respectively. Growth interruptions of 60s (b) and (d) were applied after each GaAs or InAs deposition to allow rebuilding of the QP. The QP evolution during growth is monitored by reflection high-energy electron diffraction (RHEED) where the formation of the QP can be directly observed by a change of the diffraction pattern from short, streaky rods to clear chevrons. The height of the QP (e) can be directly controlled by the number of repetitions of this sequence \cite{He2007}. We studied samples with 8 and 16 repetitions which are structurally characterized by cross-sectional transmission electron microscopy (X-TEM). Figure \ref{fig1} (f) and (g) show micrographs of a 40nm QP formed by 16 repetitions of our deposition cycle. This is about twice the height of QPs formed by eight repetitions which we found to be 23nm \cite{He2007} demonstrating precise control on a nanometer scale. The bright field image (f) is sensitive to the local strain which shows up as dark regions at the edge of the QP. The Z-contrast scanning TEM mode image (g) which is sensitive to the chemical composition shows that the indium rich QPs are laterally embedded in a InGaAs matrix (M) in which the indium content is reduced compared to the pillar region. In addition, an increased indium content at the bottom and top end of the QP forming QD like structures is also observed as indicated in Figure \ref{fig1} (e) and (g). Furthermore, the QPs are terminated by GaAs in the growth direction. These TEM investigations clearly demonstrate that QPs are dislocation-free, coherently strained, columnar nanostructures embedded in a semiconductor matrix. To quantify the indium content in the different regions of the structure we probed the chemical composition using scanned energy dispersive X-ray fluorescence (EDX) with a $\sim0.5\mathrm{nm}$ electron probe and an relative experimental error of 10\%. These are summarized in Table \ref{tab1} and directly confirm our finding from Z-contrast TEM imaging.
\begin{table}[hbt]
	\centering
		\begin{tabular}{l||c|c|c|c}
			region & QP & QD-1 & QD-2 & M \\
			 \hline\hline
			In(\%)& 43 & 48 & 45 & 10\\
		\end{tabular}\\
	\caption{Indium content measured by EDX.\label{tab1}}
\end{table}


\section{Electronic structure and optical properties}

\subsection{Theory of single particle states}
\begin{figure*}[bth]
    \centering
       \includegraphics[width=.800\textwidth]{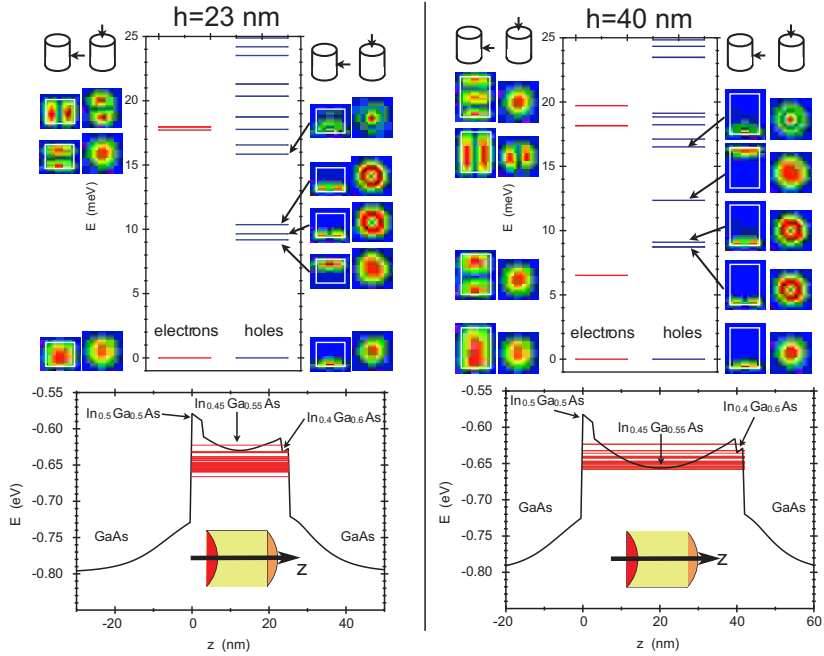}
    \caption{
    Computed single particle states for electrons and holes (upper panels) and valence band edges  (lower panels) for 23nm (left) and 40nm (right) QPs. Single particle energies are referenced to the ground state electron or hole, and the probability densities are shown in a plane containing the symmetry axis  and a plane perpendicular to the growth direction and passing through the peak in the probability density. Note that the  \textit{s}-like and  \textit{p}-like hole states are ordered differently in the two QPs because the energies are shifted differently for states localized in the top and bottom of the QPs. The valence band edges are plotted along the symmetry axis of the post, showing the energy of the highest valence state with $k=0$ in a bulk material having the same strain as the local value in the structure. The horizontal lines show the energies of the confined states.
\label{fig2}}
\end{figure*}
To gain further insight into the electronic structure of QPs we computed single particle energies and wavefunctions using an 8-band strain-dependent $k\cdot p$ model \cite{Pryor1998}.
We assumed the QPs had the composition of Table \ref{tab1} and the overall geometry shown in Fig. \ref{fig1}  with heights of 23nm and 40nm.
Material parameters were taken from reference \cite{Vurgaftmann}, including alloy bowing parameters where available.
In Fig. \ref{fig2} we show the computed single particle spectra and the corresponding probability densities from the 8-band wavefunctions.
The valence band structure is strongly influenced by the local strain and indium content, resulting in potential wells at the ends of the QP separated by a barrier. The main difference between the two QPs is a higher barrier in the 40nm QP due to the larger local strain which increases the bandgap near the middle. In both structures the holes are localized at the ends of the QP due to their large effective mass, while the lighter electrons are spread out over the entire QP. For the 23nm QP the lateral dimension is comparable to the height and hence the first and second excited electron levels are almost degenerate 18meV above the ground state.  These excited states exhibit nodes corresponding to the vertical and radial quantization reflecting the similar confinement lengths. When the QP height increases to 40nm the state corresponding to the vertical (QP axis) direction shifts downward to 7meV above the ground state while the radial-excitation state is unaffected, similar to the excited hole levels. The shift of the vertical-excitation state clearly demonstrates that the intraband transitions can be tailored by the QP height, opening a wide field of applications for mid-infrared and THz applications and devices.


\subsection{Photoluminescence}
We probed the electronic structure of QPs by low temperature ($T$=7K) micro-photoluminescence spectroscopy ($\mathrm{\mu}$-PL). Carriers were photogenerated by a Titanium-Sapphire laser focused to a $\sim1\mathrm{\mu m}$ spot. The QP emission was collected, dispersed by a 0.19m monochromator and detected by a Si-CCD camera with a spectral resolution of $\lesssim0.5$meV. QPs with low surface density and heights of 23 and 40nm were embedded in undoped GaAs or in the center of a 220nm thick intrinsic region of a n-i-p diode.
\begin{figure*}[htbp]
	\centering
		\includegraphics[width=0.5\textwidth]{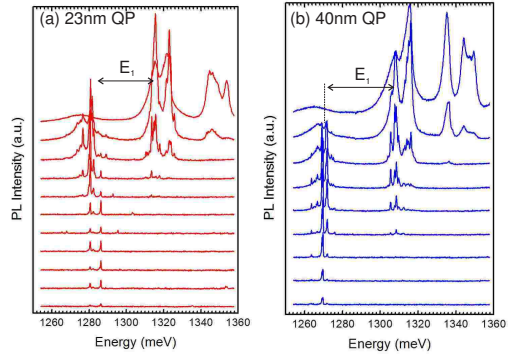}
	\caption{Power dependent PL of a single 23nm (a) and 40nm (b) Quantum Post}
	\label{fig3}
\end{figure*}

Fig. \ref{fig3} shows typical $\mathrm{\mu}$-PL spectra for a single 23nm (a) and 40nm (b) as a function of excitation density increasing by a factor of 200 from the bottom to top spectrum. We observe a clear shell-filling behavior similar to regular self-assembled QDs starting 
from a ground state emitting at 1280meV (1270meV) for the 23nm (40nm) QP. We find the first excited state emission independent of the QP height shifted by $E_1=30-40$ meV to higher energy.
In contrast to QDs this shell consists of two peaks. Remarkably, the power-dependent evolution does not change significantly when the height of the QP is increased by a factor of $\sim2$. One possible explanation for this behavior are parity selection rules which exclude transitions of the electron states which are excited in the \textit{z}-direction with the lowest hole level giving rise to a splitting $E_1$ determined by the lateral quantization, quasi-independent on the QP height. Furthermore, for the 40nm QP the splitting between the electron ground state and the first excited state is expected to be $\sim 7$meV which is smaller than typical electron-hole Coulomb energies in low-dimensional nanostructures and could make these effects not resolvable in the interband spectrum.

\subsection{Quantum Confined Stark Effect}
\begin{figure*}
	\centering
		\includegraphics[width=0.75\textwidth]{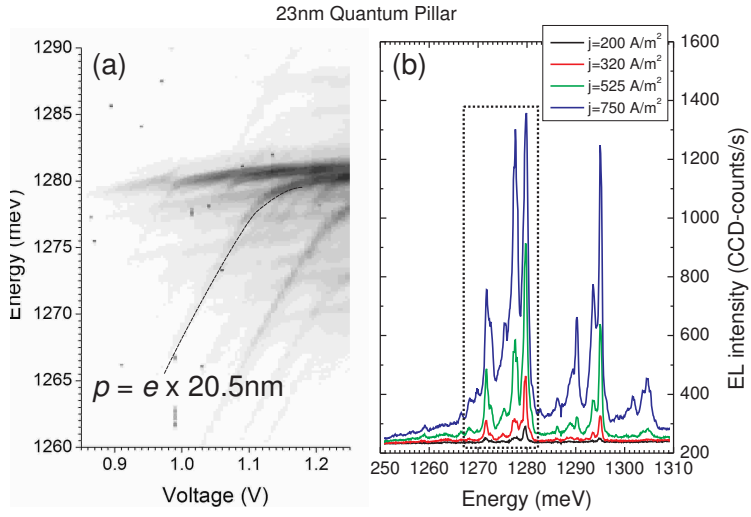}
	\caption{Bias dependent PL (a) and EL (b) of a 23nm QP. The box on (b) marks emission lines corresponding to the QP shown in (a).}
	\label{fig4}
\end{figure*}
To confirm the existence of a delocalized electron state we studied QPs embedded in n-i-p diodes. By changing the applied gate voltage $(V_B)$ we can tune the electric field ($F$) drop across the QPs. For our devices $F$ is increasing with decreasing $V_B$ with $F=0\mathrm{kV/cm}$ at $V_B\sim1.4$V. In Fig. \ref{fig4} (a) we present voltage-dependent PL spectra of 23nm QP in grayscale representation. We find lines which shift weakly as a function of $V_B$ as well as lines which exhibit a strong, linear shift. These shifts occur due to the Quantum Confined Stark Effect (QCSE) given by $\Delta E = p\cdot \Delta F$ with $p=e\cdot s$ being the static dipole moment of the exciton with an electron-hole separation $s$ \cite{Shen2002,Krenner2005,Krenner2006}. Similar to coupled QDs we observe lines which exhibit a weak Stark-shift at low $F$ (high $V_B$) which continuously changes to a strong and linear shift at higher $F$ (smaller $V_B$). This behavior can be explained as arising from a transition of a spatially 'direct' exciton where both electron and hole are localized at the same end of the QP due to their mutual Coulomb attraction to a spatially 'indirect' exciton where the two carriers are at different ends. The latter exhibit a large $p$ which is determined almost entirely by the height of the QP. Furthermore, it is independent of  $F$ ($V_B$), giving rise to a linear Stark-effect. An analysis of these slope gives $p=e \cdot 20.5$nm which is in excellent agreement with the nominal QP height of 23nm. When varying the QP height we find a systematic change of the static dipole moment \cite{Krennerunpublished}. As the exciton undergoes the transition from 'direct' to 'indirect' character its $e$-$h$ wavefunction overlap and oscillator strength vanish.\\
 For larger forward bias the carriers can be injected electrically into QPs and sharp-line electroluminescence (EL) is detected as shown in Fig. \ref{fig4} (b) as a function of the current density (\textit{j}).The group of spectral lines marked in Fig. 4(b) corresponds to the same QP probed by PL shown in (a). This observation clearly demonstrates efÞcient electrical pumping of QPs as required for optoelectronic device applications. 
 
\subsection{Outlook} 
In addition, for low $F$ we expect a regime where the exciton is fully delocalized. Here, excitons for which both the electron and hole component completely extend over the entire QP, should exhibit a large oscillator strength due to its increased coherence volume. This large oscillator strength makes QP very attractive for cavity quantum electrodynamics (cQED) experiments. Compared to regular QDs a smaller quality factor-mode volume ratio ($Q/V_{mode}$) is required \cite{Andreani1999}. This resonance should occur at lower $F$ relative to the transition to a completely 'indirect' state and depend on the relative energetic position of the hole levels in the end QDs analogous to coupled QDs \cite{Krenner2005b}. Moreover, oscillator strength and transition energies are largely tunable giving access to a whole range of experiments not possible with regular QDs. In coupled QDs one carrier usually remains localized even if the system is tunable and the maximum coherence volume and maximum oscillator strength are decreased compared to a single QD as recently confirmed experimentally \cite{Gerardot2005,Nakaoka2006}.\\

In summary, we demonstrate a method to grow height-controlled Quantum Posts. We find that the measured structural properties of the QPs have strong impact on the optical properties. For QPs embedded in n-i-p junctions we find large Stark-shifts which are in good agreement with the QP height and imply electron delocalization over the entire QP. In addition, we demonstrate sharp line EL which is required for device applications. Furthermore, a large, tunable oscillator strength is expected which makes QPs particularly attractive for cQED experiments as well as for optoelectronic and quantum information devices.\\

The authors acknowledge support through NSF via NIRT grant CCF-0507295 and Harvard-NSEC. H.J.K. thanks Alexander-von-Humboldt Foundation for financial support.

\end{document}